\documentclass[times]{qjrms4}
\pdfoutput=1

\usepackage{graphicx, color}
\usepackage{caption}
\usepackage{subcaption}
\usepackage{bm}

\newcommand{\be}{\begin{equation}}
\newcommand{\ee}{\end{equation}}
\newcommand{\bc}{\begin{center}}
\newcommand{\ec}{\end{center}}

\newcommand{\lam}{\lambda}

\def\H{{\bf{H}}}

\def\x{{\bf{x}}}
\def\y{{\bf{y}}}

\def\R{{\bf{R}}}

\def\w{{\bf{w}}}
\def\T{{\bf{T}}}
\def\X{{\bf{X}}}

\def\P{{\bf{P}}}
\def\e{{\bf{e}}}
\def\T{{\bf{T}}}
\def\S{{\bf{S}}}

\def\V{{\bf{V}}}

\def\et{{\bm{\eta}}}

\def\f{{\bf{f}}}

\def\fm{\hat{{\bf{f}}}}
\def\gm{\hat{{\bf{g}}}}
\def\lam{{\bm{\lambda}}}
\def\lamm{\hat{{\bm{\lambda}}}}

\def\xm{\hat{{\bf{x}}}}
\def\ym{\hat{{\bf{y}}}}

\def\zobs{{\bf{\y}}^o}

\def\Id{{\bf{I}}}

\def\Pf{{\bf{P}}^f}
\def\PfI{({\bf{P}}^f)^{-1}}

\def\KR{{\bf{K}}}

\def\Pm{{\bf{P}}_m}

\usepackage[colorlinks,bookmarksopen,bookmarksnumbered,citecolor=red,urlcolor=red]{hyperref}

\newcommand\BibTeX{{\rmfamily B\kern-.05em \textsc{i\kern-.025em b}\kern-.08em
T\kern-.1667em\lower.7ex\hbox{E}\kern-.125emX}}

\usepackage{moreverb}
\def\volumeyear{2014}

\usepackage{natbib}
\begin{document}

\runningheads{L. Mitchell and A. Carrassi}{Accounting for model error due to unresolved scales within ensemble Kalman filtering}


\title{Accounting for model error due to unresolved scales within ensemble Kalman filtering}

\author{Lewis Mitchell\affil{a,b}\corrauth 
\ and Alberto Carrassi\affil{c,d}}

\address{\affilnum{a}Computational Story Lab,
  Department of Mathematics \& Statistics,
  Vermont Complex Systems Center
  \& the Vermont Advanced Computing Core,
  The University of Vermont,
  Burlington, 
  VT 05401,
  USA.\\
  \affilnum{b}School of Mathematical Sciences,
  University of Adelaide,
  Adelaide,
  SA 5005,
  Australia.\\
\affilnum{c}NERSC - Nansen Environmental and Remote Sensing Center, Bergen, Norway.\\
\affilnum{d}IC3 - Institut Catala de Ciencies del Clima, Barcelona, Spain.}

\corraddr{lewis.mitchell@adelaide.edu.au}

\date{\today}

\begin{abstract}
We propose a method to account for model error due to unresolved scales in the context of the ensemble transform Kalman filter (ETKF). 
The approach extends to this class of algorithms the deterministic model error formulation recently explored for variational schemes and extended Kalman filter.
The model error statistic required in the analysis update is estimated using historical reanalysis increments and a suitable model error evolution law.     
Two different versions of the method are described; a time-constant model error treatment where the same model error statistical description is time-invariant, and a time-varying treatment where the assumed model error statistics is randomly sampled at each analysis step. 
We compare both methods with the standard method of dealing with model error through inflation and localization, and illustrate our results with numerical simulations on a low order nonlinear system exhibiting chaotic dynamics. 
The results show that the filter skill is significantly improved through the proposed model error treatments, and that both methods require far less parameter tuning than the standard approach.
 Furthermore, the proposed approach is simple to implement within a pre-existing ensemble based scheme.  
 The general implications for the use of the proposed approach in the framework of square-root filters such as the ETKF are also discussed. 
 \end{abstract} 


\maketitle


\section{Introduction}

A fundamental concern within all data assimilation schemes is the treatment of model error,
which can arise from, amongst other sources, the impossibility of characterizing phenomena at all scales of motion.
It is inevitable that there will be some processes which remain unresolved in any numerical environmental model,
making the problem of model error of fundamental importance.

Representing model error within data assimilation is an important consideration within modern-day operational data assimilation algorithms,
and there has appeared in recent years a significant body of work dedicated to this problem.
The state augmentation method \citep{Jazwinski} was primarily introduced in the context of Kalman filtering. In this method, the state estimation problem is formulated in terms of an augmented state vector which includes, along with the state estimate, a set of parameters used for the model error representation. 
This approach has been applied successfully in both Kalman filter-like and variational assimilation methods \citep{Zupanski1997, Nichols2003, Zupanski2006}. 
\cite{DeeDaSilva1998} proposed an algorithm to estimate and remove the biases in the background field in a data assimilation system due to additive systematic model errors. 
This method was implemented for the bias correction of the humidity analysis component of the Goddard Earth Observing System (GEOS) assimilation system \citep{Dee2000} and more recently in the context of the European Centre for Medium-range Weather Forecasts (ECMWF) ocean data assimilation \citep{Balmaseda2007}.
A key ingredient of the state augmentation technique is the definition of a ‘model’ for the model error (e.g. \citet{Nichols2003}).
Recently \cite{Zupanski2006} have described the model error evolution using a first-order Markov process.
A similar assumption was already used in \cite{Daley1992b} to investigate the impact of time-correlated model errors in Kalman filtering.
In \cite{Dee2005} the model bias was modelled based on unknown bias parameters to be determined and on the use of analysis increments; these latter are also used here as explained in the sequel. 

In ensemble-based data assimilation procedures, \cite{Houtekamer2009} examine both isotropic and stochastic model error representations within an operational numerical weather prediction (NWP) system,
while \cite{Charron2010} study stochastic methods within the same system.
One of the most direct methods to account for model error in ensemble filtering is through tuning the inflation and localization parameters.
These parameters,
which were originally introduced to ameliorate the covariance underestimation \citep{Anderson1999}  and spurious long-range correlations \citep{Hamill2001,Houtekamer1998} associated with finite ensemble size effects,
nonetheless allow the user some freedom to compensate for other sources of error including model error.
Covariance inflation in particular can be useful for achieving this aim,
as it has the effect of decreasing the amount by which the analysis ``trusts'' the model forecast relative to the observations.
This parameter of course requires empirical tuning, which may be computationally expensive.
Regardless, inflation has been shown to be effective approach for dealing with model error in operational models \citep{Deng2011,Raynaud2012}.

We propose a deterministic approach to the model error problem in ensemble data assimilation,
inspired by previous works on model error treatment for both variational \citep{Carrassi2010} and sequential \citep{Carrassi2011} schemes.
This method assumes that the model error is a time-correlated process in order to derive expressions for the model error bias and covariance, 
along with an approximated dynamical law suitable for practical applications.
The bias and covariance are then approximated from an assumed database of historical analysis increments.

Recognizing that there are many ways in which this information about model error may be incorporated into the data assimilation process,
we investigate two ways in which we might implement our model error treatment within an ensemble Kalman filter.
The first method treats model error as deterministic and constant in time similar to \cite{Carrassi2011},
while the second allows for the model error correction to vary in time in a random manner, 
but where the model noise is sampled from the model error statistics derived using the deterministic hypothesis.
The latter approach leads to a randomly perturbed forecast model somewhat analogous to stochastic climate models,
which have been shown to be beneficial in ensemble data assimilation schemes \citep{Harlim2008,Mitchell2012}.
The proposed approaches are studied and compared in the framework of the ensemble transform Kalman filter (ETKF) \citep{Bishop2001}, a prototype square-root filter \citep{Tippett2003}. 
The implications of this choice for the performance of the model error treatment are also discussed. 
The advantages and drawbacks of the method in relation with standard inflation/localization procedures for treating model error are highlighted with regards to their implementation in a more realistic model and observational scenario.

An outline of the remainder of the paper is as follows:
in Section 2 we formulate the problem of model error due to unresolved scales as a time-correlated process,
and derive expressions for the model error bias and covariance in the short-time regime.
In Section 3 we describe the ensemble transform Kalman filter,
and discuss strategies for dealing with model error in Section 4.
We present numerical results applying these methods to a low-dimensional nonlinear model with chaotic dynamics in Section 5,
and conclude with a discussion of our results in Section 6.

\section{Problem formulation\label{sec:formulation}}

Let the model at our disposal be represented as:
\begin{equation}
\label{eqn:model}
\frac{d\x(t)}{dt} = \f(\x, \lam),
\end{equation}
where $\f$ is typically a nonlinear function, defined in ${\mathbb R}^N$ and $\lam$ is a $P$-dimensional vector of parameters.

Model (\ref{eqn:model}) is used to describe the evolution of a (unknown) ``true" dynamics or {\it nature}, whose evolution is assumed to be given by the following coupled equations:
\begin{equation} 
\label{nature}
 \frac{d\xm (t)}{dt} = \fm(\xm,\ym,\lamm) \qquad \frac{d\ym (t)}{dt} = \gm(\xm,\ym,\lamm) 
\end{equation} 
where $\xm$ is a vector in ${\mathbb R}^N$, and $\ym$ is defined in ${\mathbb R}^L$ and may represent scales that are present in the real world, but are neglected in model (\ref{eqn:model}); the unknown parameter $\lamm$ has dimension $P'$ different from $P$. 
The true state is thus a vector of dimension $N+L$. While the model state vector $\x$ and the variable $\xm$ of the true dynamics span the same phase space, the difference in the functions $\f$ and $\fm$ implies that they do not in general have the same attractor.

When using the model (\ref{eqn:model}) to describe the evolution of $\xm$, estimation error can arise from the uncertainty in the initial conditions in the resolved scale ($\x(t_0)\ne \xm(t_0)$) and from the approximate description of the nature afforded by (\ref{eqn:model}); we refer to this type of error as {\it model error}. A number of different sources of model error are present in environmental modeling. Typical examples are those arising from the inadequate description of some physical processes, numerical discretization and/or the presence of scales in the actual dynamics that are unresolved by the model. In this study, we focus on the latter and assume furthermore that the model parameter $\lamm$ is perfectly known so that $\lam=\lamm$.

Following the approach outlined by \cite{Nicolis2004}, we derive the evolution equations for the dominant moments, the mean and covariance, of the estimation error $\delta\x = \x - \xm$ in the resolved scale ({\it i.e.} in ${\mathbb R}^N$). The formal solutions of (\ref{eqn:model}) and (\ref{nature}) read respectively:
\begin{equation}  
\label{modsol}
\x(t) = \x_0 + \int_0^t d\tau \f(\x(\tau),\lam)
\end{equation}
\begin{equation}  
\label{trusol}
\xm(t) =\xm_0 + \int_0^t d\tau \fm(\xm(\tau),\ym(\tau),\lamm)
\end{equation}
where $\x_0=\x(t_0)$, and $\xm_0 = \xm(t_0)$.
By taking the difference between (\ref{modsol}) and (\ref{trusol}), and averaging over an ensemble of perturbations around the reference trajectory, we obtain the evolution equation for the mean error, or bias:
\begin{equation}
\label{bias}
\left\langle\delta\x(t)\right\rangle = \left\langle \delta\x_0\right\rangle + \int_0^t d\tau \left\langle \delta \f (\tau) \right\rangle
\end{equation}
with  $\delta\x_0 = \x_0 - \xm_0$ and $\delta \f(t) = \f(\x(t),\lam) - \fm(\xm(t),\ym(t),\lamm)$.
With the hypothesis that the initial condition is unbiased, $\langle\delta\x_0\rangle=0$, Eq. (\ref{bias}) gives the evolution equation of the bias due to the model error, usually referred to as drift in the context of climate prediction \citep{DobRey-et-al-13}. The important factor driving the drift is the difference between the true and modeled velocity fields, $\left\langle \delta \f(\tau) \right\rangle$. We treat this difference as being correlated in time, and expand (\ref{bias}) in a Taylor time series around $t_0=0$ up to the first non-trivial order. By assuming an unbiased initial condition this expansion reads:
\begin{equation}
\label{bias_TC}
{\bf b}_m=\left\langle\delta\x(t)\right\rangle \approx \left\langle \delta \f (\tau) \right\rangle t
\end{equation}
Equation (\ref{bias_TC}) gives the evolution of the bias, ${\bf b}_m$, the drift, in the linear and short-time approximation and the subscript ``m" stands for model error-related bias.
  
Similarly, by taking the expectation of the external product of the error anomalies $\delta\x - \left\langle\delta\x\right\rangle $ with themselves, we have:
\begin{align}
\nonumber {\bf P}(t) &= \left\langle \left( \delta\x(t)\right)\left(\delta\x(t) \right)^T\right\rangle \\
\nonumber &= \left\langle \left( \delta\x_0\right)\left(\delta\x_0 \right)^T\right\rangle \\
\nonumber &\hphantom{= }+  
\left\langle\left( \delta\x_0\right) \left(\int_0^t d\tau \delta \f (\tau)\right)^T \right\rangle \\
\nonumber & \hphantom{= }+  
\left\langle\left(\int_0^t d\tau  \delta \f (\tau) \right) \left( \delta\x_0\right)^T \right\rangle \\
& \hphantom{= }+ \left\langle\int_0^t  d\tau \int_0^t d\tau' \left( \delta \f(\tau) \right) \left( \delta \f(\tau ')\right)^T \right\rangle
\label{cov}
\end{align}
Equation (\ref{cov}) describes the time evolution of the estimation error covariance in the resolved scale. 
The first term on the right hand side,
 which does not depend on time, represents the covariance of the initial error. The two following terms account for the correlation between the error in the initial condition and the model error, while the last term combines their effect on the evolution of the estimation error covariance. 
Following a standard hypothesis in most data assimilation procedures \citep{Jazwinski}, we assume that model and initial condition error do not correlate with each other so that the second and third terms on the RHS of (\ref{cov}) are set to zero.
By further assuming that the resolved scale initial condition error is zero, Eq. (\ref{cov}) becomes:
\begin{equation}
\label{cov-mod}
{\bf P}(t) = \left\langle\int_0^t d\tau \int_0^t d\tau' \left( \delta \f (\tau) \right) \left( \delta \f (\tau') \right)^T \right\rangle
\end{equation}
Equation (\ref{cov-mod}) represents the forecast error covariance in a deterministic process starting from perfect initial conditions but in presence of error due to the unresolved scale $\hat{{\bf y}}$. 
The amplitude and structure of this covariance depend on the dynamical properties of the difference between the nature and model velocity fields.
Note that the hypothesis of perfect resolved-scale initial conditions is only functional to our discussion in order to isolate the contribution to forecast error related to the unresolved scale. 
In realistic applications, $\delta\x_0\ne0$ and the use of (\ref{cov-mod}) requires some tuning;  
in Sections \ref{sec:methodology} and \ref{sec:results} we will discuss how to address this issue by means of correctly weighting the corrections due to model error.
Furthermore, 
our assumption that initial condition and model errors do not correlate is only valid for short times. 
Over longer times the model error feeds into the state-estimation error and its evolution begins to be dominated by the Lyapunov modes so that model and initial condition error become naturally correlated.
We investigate over what forecast windows the short-time approximation is numerically reasonable in Section \ref{sec:results}.

Assuming as above that these differences are correlated in time, we can expand (\ref{cov-mod}) in a Taylor series up to the first nontrivial order around the arbitrary initial time $t_0=0$, and obtain:
\begin{equation}
\label{cov_aprx}
{\bf P}_m(t) \approx \left\langle\left( \delta \f \left(0\right) \right) \left( \delta \f \left(0\right)\right)^T \right\rangle t^2 
\end{equation}
Equation (\ref{cov_aprx}) describes the short-time evolution of the estimation error covariance when the initial condition error is uncorrelated with the  model error. Note that, if the terms ${\bf f}-\hat{{\bf f}}$ are delta-correlated, as is in the case of uncorrelated model error, the short-time evolution of ${\bf P}(t)$ is bound to be linear instead of quadratic. 
This distinctive feature is relevant in data assimilation applications where model error is often assumed to be uncorrelated in time, a choice that can also reduce the computational cost, particularly in the case of variational assimilation \citep{Tremolet2006,Carrassi2010}.

\section{The ensemble transform Kalman filter (ETKF)\label{sec:ETKF}}

Let us assume that a set of $M\le N$ noisy observations of the resolved scale, $\xm$, is available at the regularly spaced discrete times $t_j=t_0+j\tau$, $j=1,2...$, with $\tau$ being the assimilation interval. The observations are stored as components of an $M$-dimensional observation vector ${\bf y}^o$, so that:
\begin{equation}
\label{obs}
{\bf y}^o(t_j)={\bf y}^o_j={\mathcal H}(\hat{{\bf x}}_j)+{\bm\epsilon}^{o}
\end{equation}
where $\xm_j=\xm(t_j)$ and ${\bm\epsilon}^{o}$ is the observation error, assumed to be Gaussian, uncorrelated in time, and with known covariance matrix ${\bf R}$. ${\mathcal H}$ is the (possibly nonlinear) observation operator which maps the model solution to the observation space, and may involve spatial interpolations (or spectral to physical space transformation in the case of spectral models) as well as transformations based on physical laws for indirect measurements \citep{Kalnay}. To simplify notation, the time dependency is removed hereafter from all vectors and matrices, which are then assumed to be evaluated at the same arbitrary time unless specified differently.  

In ensemble Kalman filters, an ensemble with $k$ members $\x_k$
\[
\X=\left[ \x_1,\x_2,\dots,\x_k \right]
\in \mathbb{R}^{N \times k}
\]
is propagated by the full nonlinear model dynamics (\ref{eqn:model}) according to
\begin{equation}
\frac{d\X}{dt} = {\bf{f}}(\X)\; ,
\quad
{\bf{f}}(\X) =\left[ {\bf f}(\x_1),{\bf f}(\x_2),\dots,{\bf f}(\x_k) \right]
\in \mathbb{R}^{N\times k} \;  \label{eqn:f}
\end{equation}
to produce a forecast ensemble $\X^f$; all members evolve using the same ${\bm\lambda}$. This ensemble is split into its mean 
\[
{{\x}^f} = \frac{1}{k}\sum_{i=1}^k\x^f_{i}=\X^f\w
\qquad {\rm{with}} \qquad
\w =
\frac{1}{k}\e
\in  \mathbb{R}^{k}
\; ,
\]
where $\e=\left[1,\dots,1\right]^T \in \mathbb{R}^{k}$, and ensemble deviation matrix
\begin{equation}
\label{devMat}
\X^{f\prime}=\X^f-{{\x}^f}\e^T=\X^f\mathbf{U} \; 
\end{equation}
with
\[
\mathbf{U} = \Id_{k}-\w\e^T 
\in \mathbb{R}^{k\times k}
\; .
\]
being the constant projection matrix and $\Id_{k}\in \mathbb{R}^{k\times k}\;$ the identity matrix. 
The defining characteristic of the ensemble Kalman filter (EnKF) \citep{Evensen}, is that the sample covariance of the ensemble deviation matrix $\X^{f\prime}$ is used to approximate the forecast error covariance matrix:
\begin{equation}
\P^f
= 
\frac{1}{k-1}\X^{f\prime}\left[\X^{f\prime}\right]^T
\in \mathbb{R}^{N\times N}\; .
\label{eqn:Pf}
\end{equation}
This shows the computational advantage of using the EnKF; instead of storing the $N \times N$ covariance matrix $\P^f$ one can instead store the $N \times k$ ensemble perturbation matrix $\X^{f\prime}$, resulting in a significant computational saving for $k \ll N$. Note however that $\P^f$ is rank-deficient for $k<N$; this is the typical situation in NWP,
where $N$ is of the order of $10^9$ and $k$ is of the order of $10^2$. 
This often leads to the problem of covariance underestimation, as the small number of ensemble members insufficiently capture the large number of degrees of freedom of the actual error. 

The analysis mean $\x^a$ which minimizes the cost function
\begin{align}
\nonumber J({\x}) &= 
 \frac{1}{2}({\x}^f-\x)^T\PfI(\x^f-\x) \\
 &+ \frac{1}{2}(\y^o-\mathcal{H}(\x))^T{\bf R}^{-1}(\y^o-\mathcal{H}(\x)) 
\label{eqn:J}
\end{align}
is:
\begin{equation}
\label{eqn:zaOI}
{{\x}}^a 
= 
{{\x}}^f 
+ \KR\left[\zobs - \H{{\x}}^f  \right],
\end{equation}
where
\begin{align}
\KR &= \P^f \H^T\left( \H \P^f\H^T + {\bf R}\right)^{-1}
\label{eqn:KROI}
\end{align}
is the \emph{Kalman gain matrix} or optimal (in the sense of a minimum variance estimate) weight matrix for the observations. 
The analysis error covariance matrix can be found as:
\begin{equation}
\P^a = \left(\Id_{N} - \KR\H\right) \P^f\; 
\label{eqn:PaOI}
\end{equation}
with ${\bf H}\in \mathbb{R}^{M\times N}\;$ being the linearized observation operator, and $\Id_{N} \in \mathbb{R}^{N\times N}\;$ the identity matrix.
To determine an ensemble $\X^a$ which is consistent with the analysis error covariance matrix $\P^a$ as defined in (\ref{eqn:PaOI}) and which satisfies:
\begin{equation}
\label{eqn:PaOI-2}
\P^a
= 
\frac{1}{k-1}\X^{a\prime}\left[\X^{a\prime}\right]^T\; ,
\end{equation}
where the prime denotes the deviations from the analysis mean, one can use the method of randomly perturbed observations \citep{Burgers1998,Houtekamer1998,Evensen} or a deterministic method such as an ensemble square root filter \citep{Simon}. We will use the ETKF as proposed in \cite{Bishop2001,Tippett2003,Wang2004}, which seeks a transformation $\T \in \mathbb{R}^{k\times k}$ to be applied to $\X^{f\prime}$ so that: 
\begin{equation}
\X^{a\prime}=\X^{f\prime} \T \; .
\label{eqn:TRNSF}
\end{equation}
This method has the advantage of transforming the problem into ensemble space, greatly reducing the number of operations required when $k \ll N$ as is typically the case in operational data assimilation for environmental science. 
Note that the matrix $\T$ is not uniquely determined for $k<N$, and the specific choice characterizes the type of square-root filter adopted \citep{Tippett2003}. According to \cite{Wang2004}, the ETKF transformation matrix $\T$ can be obtained by:
\begin{equation}
\label{eqn:T}
\T = 
{\bar{\V}}
\left(\Id_{k-1} + {\bar{\S}} \right)^{-\frac{1}{2}} 
{\bar{\V}}^T\;
\end{equation}
where ${\bf{U}}\S\V^T$ is the singular value decomposition of
\begin{equation}
{\bf{W}} 
= \frac{1}{k-1}
{\X^{f\prime}}^T
\H^T{\bf R}^{-1}\H
\X^{f\prime}\; .
\label{eqn:U_ETKF}
\end{equation}
The matrix ${\bar{\V}}\in \mathbb{R}^{k\times (k-1)}$ is obtained by erasing the last zero column from $\V\in\mathbb{R}^{k\times k}$, and ${\bar{\S}} \in\mathbb{R}^{(k-1)\times (k-1)}$ is the upper left $(k-1)\times(k-1)$ block of the diagonal matrix $\S\in\mathbb{R}^{k\times k}$. The presence of the null eigenvalue is due to the fact that the $k$ ensemble deviation are not independent in view of (\ref{devMat}). 
The transformation (\ref{eqn:T}) is mean preserving, so that $\T\e = {\bf 0}$; 
a condition which is not necessarily true for general square root filters \citep{Wang2004}.

\section{Strategies for dealing with model error\label{sec:methodology}}

The formulation of the ETKF described in Section \ref{sec:ETKF} does not explicitly include a treatment of the model error in the state estimation procedure.
This is the topic of the present Section. We first review a common approach used to deal with model error in most ensemble-based schemes including the ETKF, and then introduce two novel methods based on the deterministic formulation outlined in Section \ref{sec:formulation}.

\subsection{ETKF with inflation and localization}

The standard way to account for model error in most ensemble-based data assimilation schemes is through artificial inflation and localization of the error covariances. 
These solutions are often adopted with the general purpose of mitigating the misrepresentation of the actual error covariance due to the use of a finite size ensemble, 
even in the absence of model error.

Two main types of inflation procedures are known, referred to as multiplicative \citep{Anderson1999} and additive \citep{Hamill2005}.
In the former case, the matrix $\P^f$ is multiplied by a scalar coefficient, $(1+\delta)$, with $\delta$ slightly larger than zero in most relevant applications. By doing so, the variance explained by $\P^f$ is increased while its range and rank are left unchanged. The increase in the explained variance is taken to account for the portion of forecast error due to model error. Optimization of $\delta$ is usually done with numerical tuning, but some advanced solutions allowing for an adaptive estimation have recently appeared \citep{Sacher2008,Whitaker2012} as well as a new ensemble-based scheme without the intrinsic need for inflation \citep{Bocquet2011}. 
That the span of $\P^f$ is preserved makes multiplicative inflation attractive in applications with highly chaotic systems,
 where the ensemble is successful in tracking instabilities that grow along the data assimilation cycle and the range of $\P^f$ has a significant projection on the unstable subspace \citep{Palatella2013}.  
In additive inflation, random noise is added along the diagonal of $\P^f$ or $\P^a$ at each analysis step. 
This procedure refreshes the covariance matrix,
 preventing the collapse of the ensemble along few dominant modes, 
 and maintaining diversity amongst the ensemble members. 
 However,
  this gain in ensemble spread is obtained at the price of breaking the dynamical consistency of the ensemble and may cause imbalances. 
  Finally, a key issue is the prescription of the statistical properties of the noise, 
  which are difficult to determine and whose knowledge reflect assumptions made on the properties of the model error. 
  Note that the deterministic approach described in the following implicitly addresses this issue and proposes a way to describe the model error mean and covariance. 

The localization procedure is designed to remove long-range spatial correlations that, 
due to the finite size of the ensemble, 
are often poorly estimated and may introduce spurious correlations into the assimilation cycle \citep{Houtekamer1998,Hamill2001a}. 
The standard method for implementing localization is through the Schur product of $\P^f$ by a localization matrix whose entries are obtained on the basis of a predefined length-correlation function $\mathbf{\Omega}(r)$, $r$ being the distance between model grid points. Recently \cite{Bishop2011} have proposed a flow dependent localization method.    

In the numerical experiments that follow the ETKF is implemented with the simultaneous use of multiplicative inflation and localization. 
Before being used in the analysis update (\ref{eqn:J})-(\ref{eqn:KROI}),
 the forecast error covariance matrix is transformed according to: 
\begin{equation}
\label{ETKF_INLOC}
\P^f \Longrightarrow (1+\delta)\P^f\circ\mathbf{\Omega}(r).
\end{equation}
The update step is then completed with the transformation of the ensemble of forecast deviations into the analysis ensemble using (\ref{eqn:TRNSF}).

\subsection{ETKF with deterministic model error treatment}

This approach is inspired by previous studies on deterministic model error treatment in the context of both variational and sequential schemes \citep{Carrassi2010, Carrassi2011}.   
Model error is regarded as a time-correlated process and the short-time evolution laws (\ref{bias_TC}) and (\ref{cov_aprx}) are used to estimate the bias, ${\bf b}_m$, and the model error covariance matrix, $\Pm$, respectively. 
The sequential nature of the ETKF justifies adoption of the short-time approximation, 
however an important practical concern is the ratio between the duration of the short-time regime and the length of the assimilation interval $\tau$ over which the approximation is used \citep{Nicolis2004}. 

A key issue is the estimation of the first two statistical moments of the advection mismatch, $\f -\fm$, required in (\ref{bias_TC}) and in (\ref{cov_aprx}) respectively. 
This problem is addressed here assuming that a reanalysis dataset of relevant geophysical fields is available and is used as a proxy of the true nature evolution. Reanalysis programs constitute the best-possible estimate of the Earth system over an extended period of time, using an homogeneous model and data assimilation procedure, and are of paramount importance in climate diagnosis  (see \cite{Dee2011}). 

Let us suppose to have access to such a reanalysis which includes the analysis, ${\bf x}^a_r$, and the forecast field, ${\bf x}^f_r$, so that ${\bf x}^f_r(t_j+\tau_r) = \mathcal{M}({\bf x}^a_r (t_{j}))$, and $\tau_r$ is the assimilation interval of the data assimilation scheme used to produce the reanalysis. 
The operator $\mathcal{M}$ represents the forward model propagator, relative to the model (\ref{eqn:model}), over the reanalysis assimilation interval $\tau_r$; the subscript $r$ stands for reanalysis. 
Under this assumption the following approximation is made:
\begin{align}
\nonumber{\bf f}({\bf x},\lambda) &- \hat{{\bf f}}(\hat{{\bf x}},\hat{{\bf y}},\lambda) \\
\nonumber&= \frac{d{\bf x}}{dt} - \frac{d\hat{{\bf x}}}{dt} \\
\nonumber&\approx
\frac{{\bf x}^f_r(t+\tau_r) - {\bf x}^a_r(t) }{\tau_r} - \frac{{\bf x}^a_r(t+\tau_r) - {\bf x}^a_r(t) }{\tau_r} \\
\nonumber&= \frac{{\bf x}^f_r(t+\tau_r) - {\bf x}^a_r(t+\tau_r) }{\tau_r} \\
&= - \frac{\delta{\bf x}^a_r}{\tau_r}
\label{deriv_apx}
\end{align}
The difference between the analysis and the forecast, $\delta{\bf x}^a_r$, is usually referred to in the data assimilation literature as an {\it analysis increment}.
From (\ref{deriv_apx}) we see that the vector of analysis increments can be used to estimate the difference between the model and the true nature.
A similar approach was originally introduced by \cite{Leith1978}, and has been used recently to account for model error in data assimilation \citep{Li2009}.
\cite{Rodwell2007} also used a similar approach of averaging analysis increments to represent the ``systematic forecast tendencies'' of a model and diagnosed differences between its climate and the true climate. 

Note that the estimate (\ref{deriv_apx}) neglects the analysis error, so that its accuracy is connected to that of the data assimilation algorithm used to produce the reanalysis.
 This in turn is related to characteristics of the observational network such as the number, distribution and frequency of the observations. 
However, this analysis error is present and acts as an initial condition error, a contribution which is already partially accounted for in the ETKF update by the ensemble-based forecast error covariance, $\P^f$. 
As a consequence, when (\ref{cov_aprx}) is used to estimate only the model error component, an overestimation is expected. This effect, that was also found by \cite{Carrassi2011} in the context of the extended Kalman filter, can be overcome by an optimal tuning of the amplitude of ${\bf b}_m$ and ${\bf P}_m$.

Substituting (\ref{deriv_apx}) into (\ref{bias_TC}) we obtain an approximation of the bias at the analysis time, ({\it i.e.} the drift between $t$ and $t+\tau$), as:
\begin{equation}
\label{bias_TC-2}
\overline{{\bf b}}_m = \left\langle\delta\x^a_r\right\rangle\frac{\tau}{\tau_r}
\end{equation}
Inserting (\ref{deriv_apx}) into (\ref{cov_aprx}), the model error contribution to the forecast error covariance can be estimated taking the external product of (\ref{deriv_apx}) after removing the mean and reads:
\begin{equation}
\label{cov_aprx_3}
\overline{\P}_m = \left\langle\left(\delta{\bf x}^a_r - \left\langle\delta{\bf x}^a_r\right\rangle \right) \left(\delta{\bf x}^a_r - \left\langle\delta{\bf x}^a_r\right\rangle    \right)^T \right\rangle \frac{\tau^2}{\tau_r^2} 
\end{equation}

\subsubsection{ETKF with time-constant model error treatment - {\bf ETKF-TC}}

In this approach, ${\bf b}_m$ and $\Pm$ are estimated once using (\ref{bias_TC-2}) and (\ref{cov_aprx_3}) and then kept constant along the entire assimilation cycle. 

In the ETKF-TC the ensemble forecast mean and covariance are transformed according to:
\begin{equation}
\label{ETKFTC_b}
\x^f \Longrightarrow \x^f - \alpha\bar{{\bf b}}_m,
\end{equation}
\begin{equation}
\label{ETKFTC_P}
\P^f \Longrightarrow (1+\delta)\P^f\circ\mathbf{\Omega}(r) + \alpha^2\overline{\P}_m
\end{equation}
The inflation and localization apply to the ensemble-based forecast error covariance matrix $\P^f$ alone.
The scalar term $\alpha>0$ is a tunable coefficient aimed at optimizing the bias size to account for the expected overestimation connected with the use of (\ref{deriv_apx}).
The new first guess and forecast error covariance, (\ref{ETKFTC_b}) and (\ref{ETKFTC_P}), are used in the ETKF analysis formulas (\ref{eqn:J})-(\ref{eqn:KROI}) to obtain the analysis.
The update of the ensemble members is done as in the standard ETKF transforming the analysis deviation ensemble into the forecast ones according to (\ref{eqn:TRNSF}). 
In the ETKF-TC model error is repeatedly corrected in the subspace spanned by the range of $\overline{\P}_m$ where model error is supposed to be confined.
This choice reflects the assumption that the impact of the unresolved scales on the forecast error does not fluctuate greatly along the analysis cycle.

\subsubsection{ETKF with time-varying model error treatment - {\bf ETKF-TV}}

This is a modification of the ETKF-TC in which the model error correction is made time-varying by incorporating a random sampling from the assumed model error statistics into the forecast step.
A model error vector is added to each ensemble forecast member, before being used in the analysis update. 
In the ETKF-TV the forward propagation of the ensemble members, which is done in the ETKF using the model according to (\ref{eqn:f}), is substituted by:

\begin{align} 
\label{TV}
\x^{f}_{i,j} &= \mathcal{M}(\x_{i,j}^{a}) - \alpha\et_{i,j}\frac{\tau}{\tau^r}\\
\nonumber\text{with }\et_{i,j} &\in \mathcal{N}(\overline{{\bf b}}_m,\overline{\P}_m),\quad i=1,...,k 
\end{align} 

where $j$ refers to the arbitrary analysis time $t_j$.
The model error vectors, $\et_{i,j} \in \mathbb{R}^{N}$, are sampled from a distribution of the model error statistics, with mean and covariance equal to $\overline{{\bf b}}_m$ and $\overline{\P}_m$ calculated using (\ref{bias_TC-2}) and (\ref{cov_aprx_3}).
The ETKF-TV preserves the assumed model error statistics while exploring a different realization of the model error at each analysis time. 
We remark that adding a random perturbation ameliorates problems due to rank deficiency of the finite size ensemble, 
but can upset balance and introduce spurious variability.
This formulation bears a formal resemblance to stochastic climate models \citep{Majda2001},
which have recently been found to be beneficial in the context of data assimilation \citep{Harlim2008,Mitchell2012}.
Whereas these stochastic climate models rely on the mathematical technique of homogenization which may be difficult or impossible to apply to large models,
the ETKF-TV only requires access to an historical reanalysis dataset to generate $\overline{\P}_m$.
While therefore less mathematically rigorous than stochastic climate model methods,
the ETKF-TV may nonetheless provide a tractable computational approach for dealing with model error in data assimilation systems with large geophysical models.

The forecast ensemble members (\ref{TV}) are then used to build the forecast error covariance matrix $\Pf$ according to (\ref{eqn:Pf}) as in the standard ETKF and, if desired, inflation and localization are applied using (\ref{ETKF_INLOC}).   
The analysis update (\ref{eqn:J})-(\ref{eqn:KROI}), including the forecast-analysis deviations transformation, (\ref{eqn:TRNSF}), is done as in the ETKF.

\subsection{Comments on the forecast-analysis ensemble transformation}
\label{sec:inconsistency}

The ETKF forecast-analysis transformation, ${\bf T}$, in (\ref{eqn:TRNSF}), is obtained under the requirements \citep{Wang2004} that the ensemble analysis error covariance $\P^a$ given by (\ref{eqn:PaOI-2}) is equal to the one predicted by the least-square update (\ref{eqn:PaOI}), and that the analysis ensemble is centered around the mean analysis field, ${\bf x}^a$, obtained via (\ref{eqn:J})-(\ref{eqn:KROI}).
As mentioned in Section 3, when $k<N$ this transformation, which accounts for the search of a square-root of the matrix ${\bf P}^a$,
 is not unique and its choice characterizes the specific formulation of the square-root implementation \citep{Tippett2003}. 
A distinguishing feature of the ETKF is that the analysis ensemble members are orthonormal in the observation space \citep{Wang2003}. The search for a square-root of ${\bf P}^a$ is solved for the transformation matrix, which is then applied to the forecast ensemble.
The ensemble transformation implies only the rescaling and rotation of the forecast members, without introducing additional noise. 
The latter feature of the deterministic filters with respect to their stochastic counterpart causes major differences in their respective performance when applied to chaotic systems,
 particularly with relation to the ensemble size \citep{LawsonHansen2004,SakovOke2008,CVZZ2009}.      

Let us consider now the standard ETKF as given in Section 3, with inflation and localization as described in Section 4.1, so that the ensemble based forecast error covariance, $\P^f$, entering the analysis update, (\ref{eqn:J})-(\ref{eqn:PaOI}), is now given by (\ref{ETKF_INLOC}): the analysis field and the predicted error covariance are obtained assuming the inflated/localized forecast error covariance. 
Nevertheless the transformation to obtain analysis deviations consistent with this predicted $\P^a$ is applied to the forecast perturbations, $\X^{f\prime}$, evolved over the previous analysis interval. 
The fact that the inflated/localized matrix, $(1+\delta)\P^f\circ\mathbf{\Omega}(r)$, is used in the analysis update of the mean state, while the unaltered forecast deviations are used in the analysis error covariance update, causes an inconsistency: the matrix $\X^{f\prime}$ is not a square-root of $(1+\delta)\P^f\circ\mathbf{\Omega}(r)$.  
Note that we have restricted our discussion to the standard ETKF with inflation/localization only for the sake of clarity; the same arguments holds for the ETKF-TC, to an extent related to the properties of $\overline{\P}_m$. This issue arises in all circumstances when the ensemble-based forecast error covariance of a square-root filter is adjusted to account for errors misrepresented by the ensemble alone: the new matrix $\P^f$ used to update the mean, is no longer the square of $\X^{f\prime}$, and $\X^{a\prime}=\X^{f\prime}{\bf T}$ is not the square-root of the predicted $\P^a$.
This problem is not present in  the ETKF-TV, unless localization and/or inflation is applied.
In fact the model error treatment in this case acts directly on the ensemble members used to build the forecast error covariance matrix whose deviations give the corresponding square-root matrix, 
and represents a potential advantage of the ETKF-TV.
The problem of obtaining a consistent mean and ensemble deviations update was originally recognized by \cite{Whitaker2002},
while \cite{Tippett2003} outlined a general method for incorporating model error statistics into an ensemble square root filter. 

\section{Results\label{sec:results}}

To compare how well the ETKF-TC and ETKF-TV perform when compared with a standard ETKF,
we perform twin data assimilation experiments where a model trajectory taken to be the truth has noise added to create synthetic observations.
As a testbed for our experiments we take as our truth the slow-fast Lorenz-96 model \citep{Lorenz1996},
a prototype model for atmospheric dynamics consisting of a ring of $N$ large-scale, slow oscillators $\hat{x}_i$ coupled to a ring of $L=N\times J$ small-scale, fast oscillators $\hat{y}_{j,i}$:
\begin{align}
\label{eqn:L96slow} \frac{d \hat{x}_i}{d t} &= \left(\hat{x}_{i+1} +\hat{x}_{i-2}\right)\hat{x}_{i-1} - \hat{x}_i + F - \frac{hc}{b}\sum_{j=1}^{J}\hat{y}_{j,i},\\
\label{eqn:L96fast}	\frac{d \hat{y}_{j,i}}{d t} &= cb\left(\hat{y}_{j-1,i} -\hat{y}_{j+2,i}\right)\hat{y}_{j+1,i}-c\hat{y}_{j,i}+\frac{hc}{b}\hat{x}_i,
\end{align}
with $\hat{x}_{i\pm N} = \hat{x}_i$, $\hat{y}_{j\pm J,i} = \hat{y}_{j,k\pm 1}$ and $\hat{y}_{j,i\pm N}=\hat{y}_{j,i}$.

\begin{figure}
\centering
\includegraphics[width=\columnwidth]{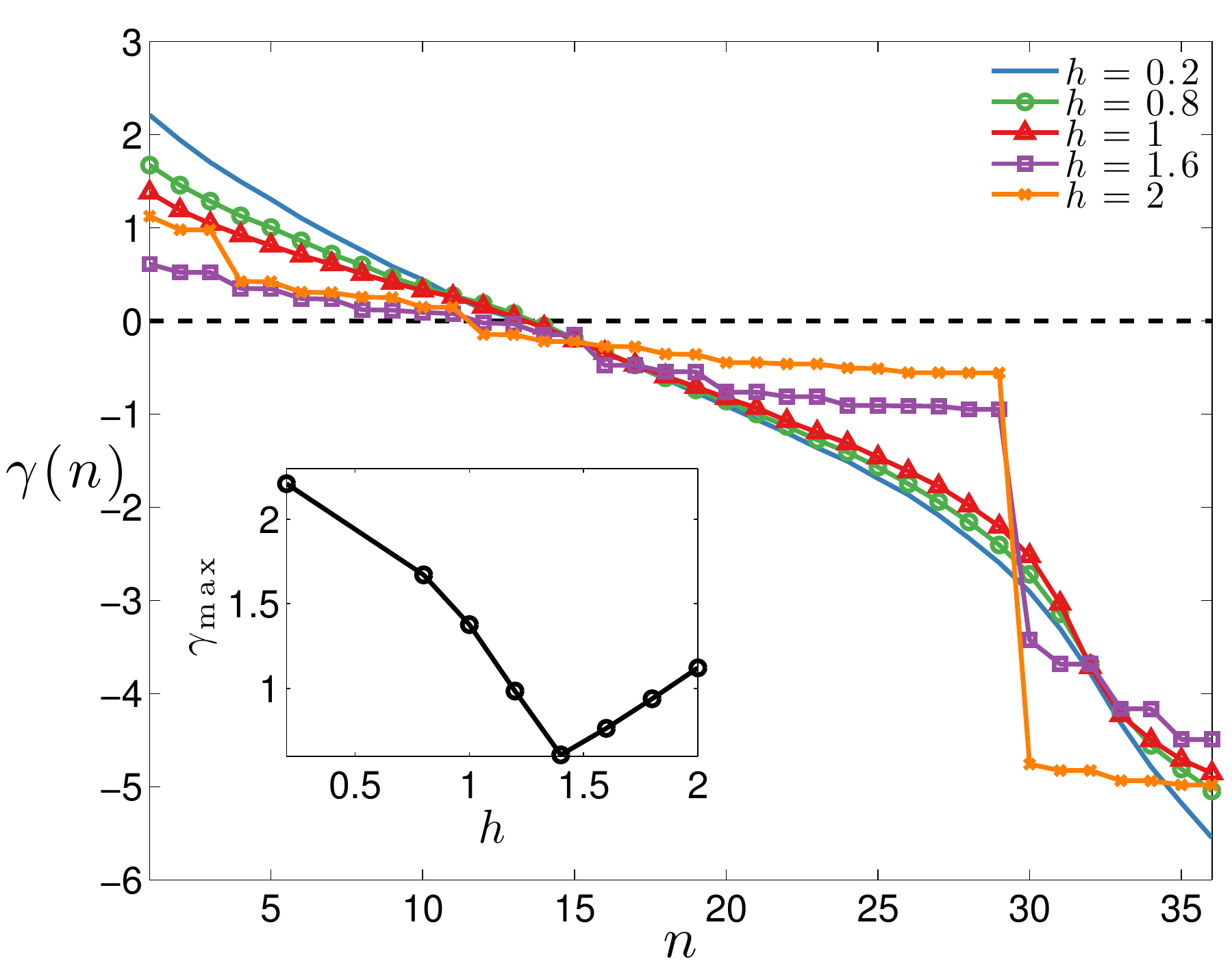}
\caption{Lyapunov spectra for a range of values of $h$. The inset shows the maximal Lyapunov exponent $\gamma_{\rm max}$, as a function of $h$.}
\label{fig:LE_h}
\end{figure}

Our model consists of the first four terms of (\ref{eqn:L96slow}) for the slow variables only:
\begin{equation}
\label{eqn:L96} \frac{d x_i}{d t} = \left(x_{i+1} +x_{i-2}\right)x_{i-1} - x_i + F
\end{equation}
with $x_{i\pm N} = x_i$, which define the resolved scale and the model has no knowledge of the fast subsystem dynamics.
The variables $\hat{x}_i$, $\hat{y}_{j,i}$ and $x_i$ refer to grid point values with specified spacing on a periodic domain,
enabling the use of localization in the following numerical experiments.
In all experiments we fix $N=36$, $J=10$, $F=10$, and $c = b = 10$ so that the fast modes oscillate 10 times faster than the slow modes.
We integrate the model using a fourth order Runge-Kutta scheme with timestep $dt = 0.005$.
For these parameters and with $h = 1$,
the climatological standard deviation is $\sigma_{\rm clim} = 3.54$
and maximal Lyapunov exponent is $\gamma_{\rm max} =  1.3775$.
In Figure \ref{fig:LE_h} we show the Lyapunov spectra of (\ref{eqn:L96slow}) for a few different values of $h$.
These spectra were calculated using a QR decomposition \citep{Parker1989} over $10^6$ time steps.
The inset shows the maximal Lyapunov exponent, 
$\gamma_{\rm max}$, 
as a function of $h$.
The maximal Lyapunov exponent decreases with increasing $h$ up to $h=1.4$, 
whereafter it increases again.
As $h$ increases past 2 the system eventually becomes non-chaotic, 
with all Lyapunov exponents $\gamma < 0$.
A more in-depth exploration of the dynamical properties of the Lorenz-96 system can be found in \cite{Frank2014}.

In all experiments we take $12$ equally spaced observations with observational error covariance $\R = 0.5 \Id$ 
or approximately $4\%$ of the climatological variance.
It is assumed that the observation error covariance matrix ${\bf R}$ is perfectly known and is used to sample the simulated observational noise. 
The use of a diagonal matrix ${\bf R}$ reflects the assumption of mutually uncorrelated observations. 
Except where otherwise stated,
we use $k= 2N = 72$ ensemble members for the ETKF analysis.

Our experiments consist of two stages: 
a {\it reanalysis} stage where the model error biases (\ref{bias_TC-2}) and covariance matrices (\ref{cov_aprx_3}) are estimated from the analysis increments $\delta \x_r^a$ of a suitably optimized ETKF;
and an {\it experiment} stage where this $\overline{\P}_m$ is used in an Observing System Simulation Experiment (OSSE),
to compare analyses made by the ETKF-TC and ETKF-TV with the standard ETKF.

\begin{figure*}
\centering
\includegraphics[width=2\columnwidth]{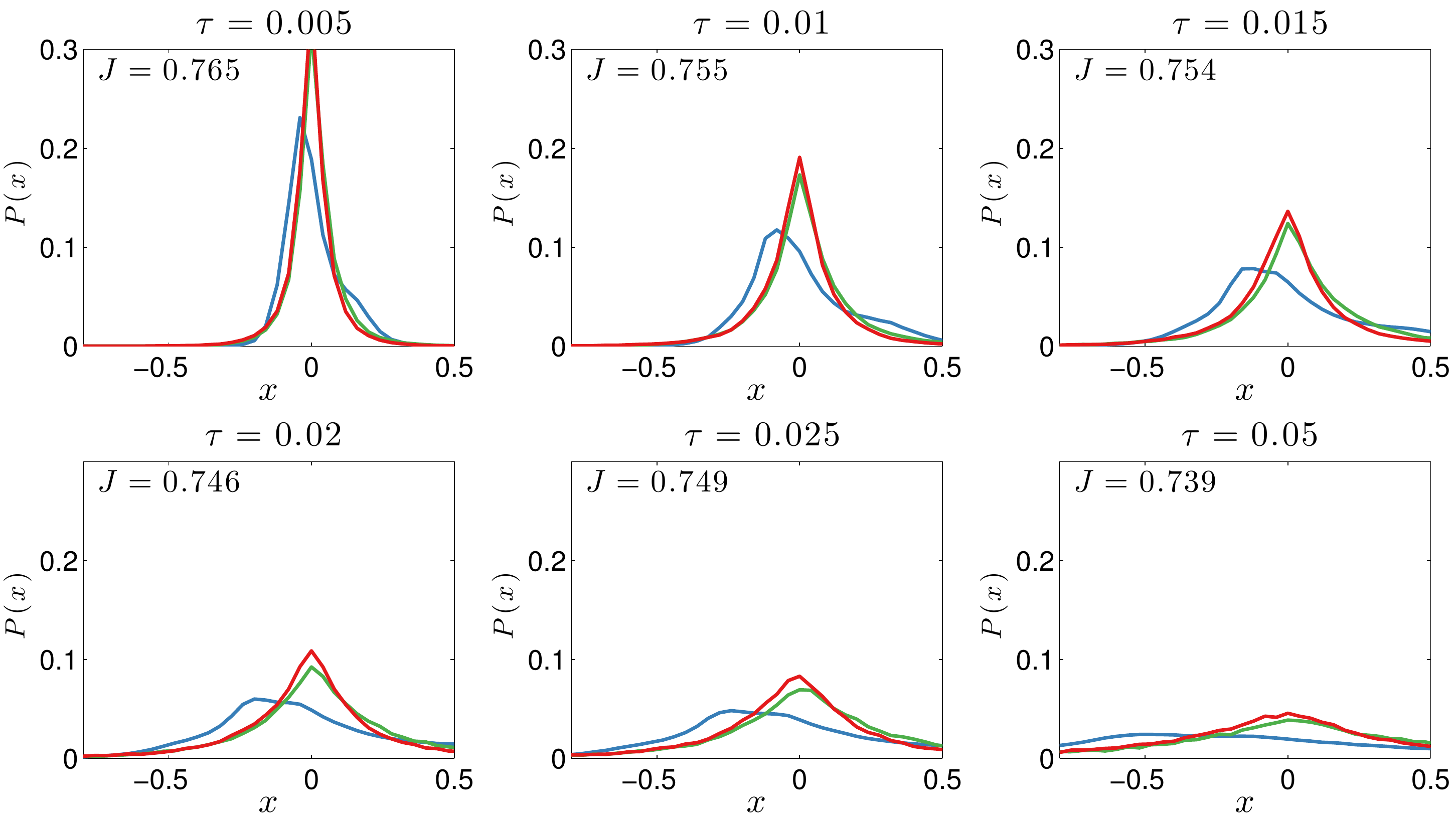}
\caption{Distribution of analysis increments for arbitrary selected observed (green) and unobserved (red) components of the Lorenz-96 system with $h=1$, as well as the ``true'' model error distribution (blue), for different forecast lengths $\tau$. 
The inset $J$-values indicate the intersection or joint probability between the true model error distribution and unobserved analysis increments.
Statistics were collated over 10 years of analyses,
and we have taken $\tau_r = \tau$ in all experiments.
}
\label{fig:pdf_dtf}
\end{figure*}

We show estimates of the model error distribution in Figure \ref{fig:pdf_dtf} for different values of the forecast interval $\tau$,
and with $h=1$.
Note that we have taken $\tau_r = \tau$ in all experiments,
meaning that the reanalysis interval is always the same as the forecast horizon of our verification. 
The blue curve in each panel shows histograms of the ``true'' model error distribution obtained by making forecasts using the model (\ref{eqn:L96}) initialized with $x$-values from one 10-year long nature trajectory of (\ref{eqn:L96slow})-(\ref{eqn:L96fast}) and considering the differences.
The green and red curves show approximations to this model error distribution arising from the analysis increments (\ref{deriv_apx}) obtained from a standard ETKF without inflation or localization for an arbitrary selected observed and unobserved variable, respectively.
The inset $J$-values indicate the intersection or joint probability between the true model error distribution and unobserved analysis increments.

Reflecting the progressive degradation of the information for increasing time intervals, both the true and approximate distributions become wider as forecasts are made longer, 
and the distributions move further apart from one another. 
This gives rise to the appearance of a bias in the distributions,
 a direct consequence of neglecting a scale of motion in our model. 
 Note that this drift would not have appeared if model error acted as a white noise. 
 Figure \ref{fig:pdf_dtf} shows also that the bias is not captured by our approximation based on the reanalysis increment. 
 This is because the ETKF, similarly to most Kalman filter-like methods, assumes the forecast and the observations are unbiased. 
 Such schemes are unable to correct this error unless a bias correction procedure is explicitly incorporated \citep{DeeDaSilva1998}.

\begin{figure*}
\centering
\includegraphics[width=2\columnwidth]{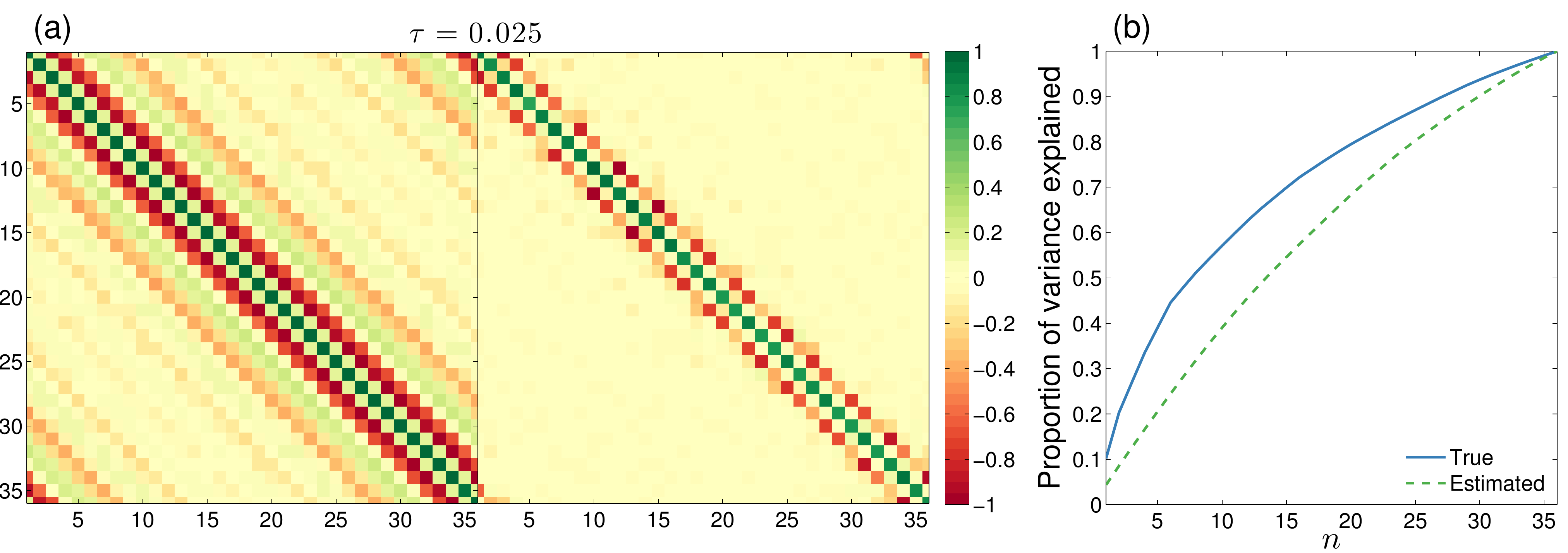}
\caption{(a): True (left) and estimated (right) model error covariance matrices $\overline{\mathbf{P}}_m$ for $\tau = 0.025$. Both matrices are normalized such that all values lie between $-1$ and $1$.
(b): Proportion of the variance explained as a function of eigenvalue number for the true and estimated covariances.
}
\label{fig:cov_dtf}
\end{figure*}

In Figure \ref{fig:cov_dtf} we show the true model error covariances (left panels) and analysis error-estimated model error covariances $\overline{\mathbf{P}}_m$ (right panels) for the same 10-year experiments as in Figure \ref{fig:pdf_dtf}.
Each heat map is normalized such that values range between $-1$ and $1$, with large-magnitude values indicating strong positive or negative correlation between sites.
The estimated model error covariances $\overline{\mathbf{P}}_m$ capture some of the negative correlations that exist between neighboring sites,
but represent little of the long-range correlation structures present in the true covariance matrix.
Correlations between pairs of observed variables are also better preserved than those between pairs involving unobserved variables.

Based on the above we set the analysis interval $\tau = 0.025$,
approximately equivalent to 3 hours,
for all subsequent experiments. 
This is as a compromise to make our short-time approximation work adequately and for the observation interval to be similar to that of a realistic observational network. 

\begin{figure}[h]
\centering
\includegraphics[width=\columnwidth]{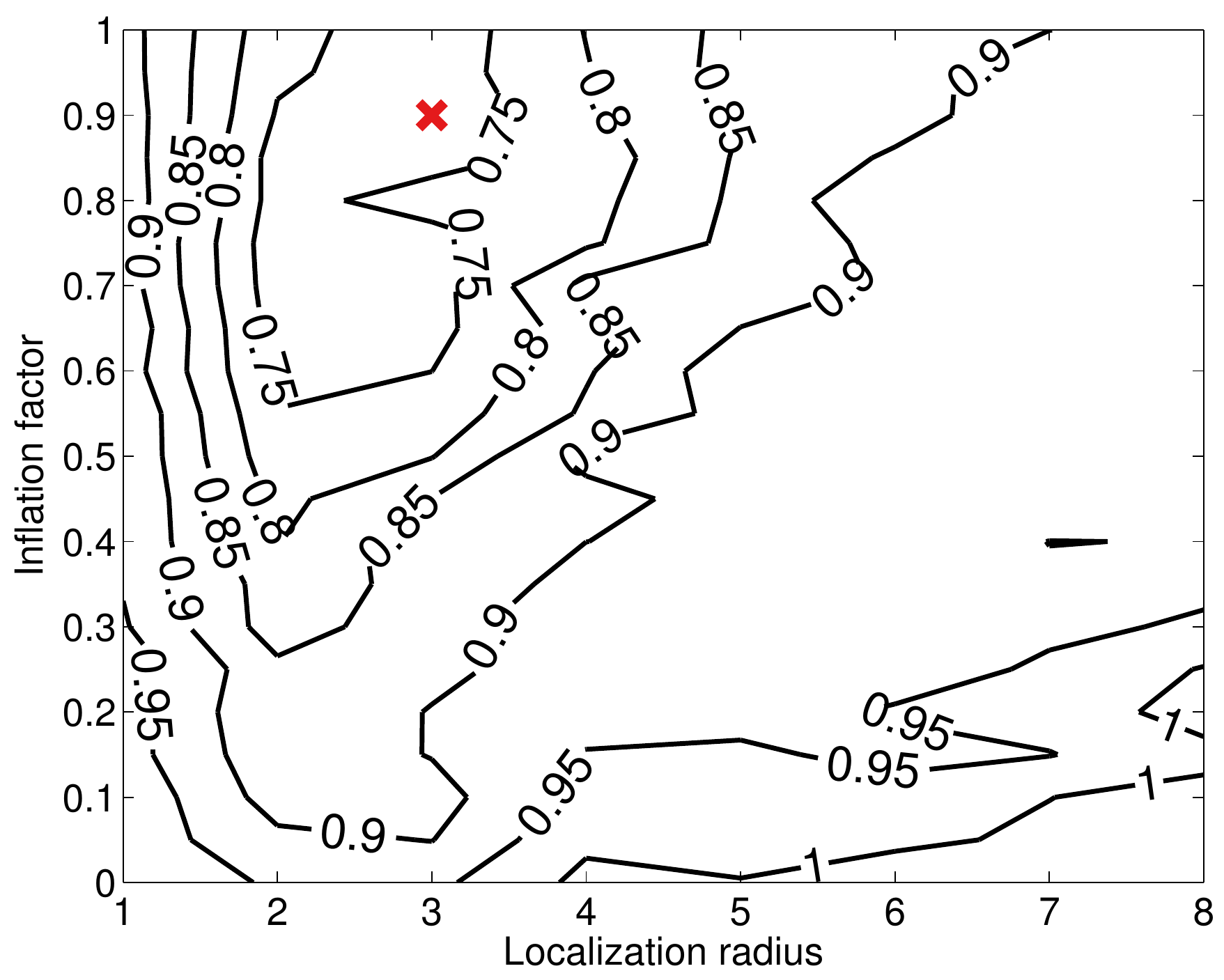}
\caption{Contour plot of RMS errors for the ETKF as a function of inflation factor $\delta$ and localization radius $r$. The red cross shows the location of the minimum. RMS errors are normalized by the climatological standard deviation $\sigma_{\rm clim}$, and are averaged over 200 realizations of one month in duration.}
\label{fig:ETKFerrAll}
\end{figure}

We first calibrate the inflation factor $\delta$ and localization radius $r$ in order to produce the best possible reanalysis data set.
Figure \ref{fig:ETKFerrAll} shows a contour plot of RMS analysis errors for a standard ETKF,
averaged over 200 realizations of one month in length. 
The RMS errors are normalized by the climatological standard deviation $\sigma_{\rm clim}$.
The best results are obtained with a strong inflation, $\delta=0.9$, and a moderate localization radius, $r=3$. 
This suggests that the ETKF tends to underestimate the actual error variance and misrepresent the error spatial correlations at a length scale larger than 3 grid-points. Based on the results in Figure \ref{fig:ETKFerrAll} we set $r=3$, $\delta=0.9$ in all subsequent experiments to ensure both accuracy and stability in the ETKF.

\begin{figure}[h]
\centering
\includegraphics[width=\columnwidth]{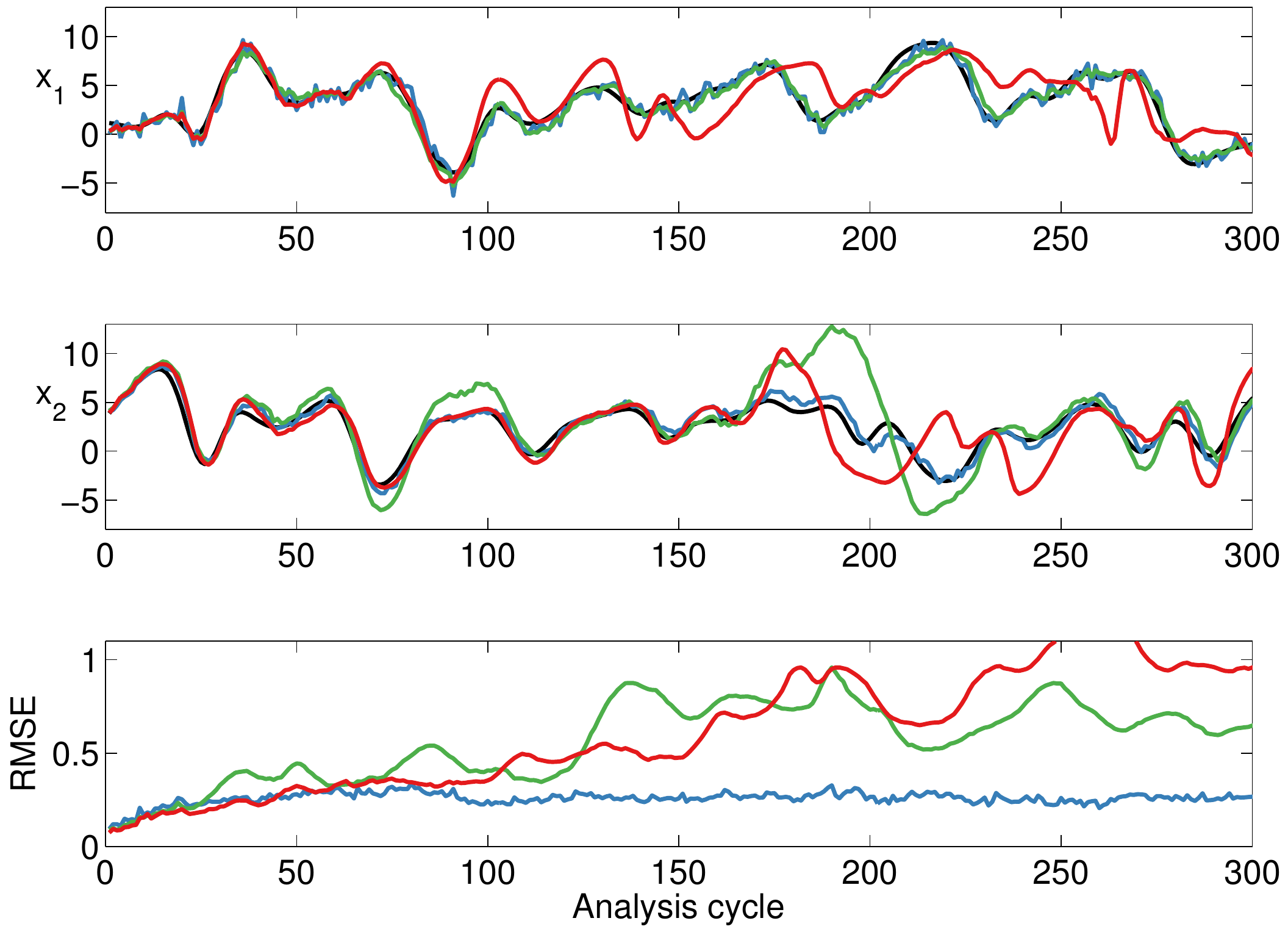}
\caption{Time series of analyses made by the ETKF (red),
ETKF-TC (green) and ETKF-TV (blue)
for an observed variable (top panel),
unobserved variable (middle panel),
and RMS errors (lower panel).}
\label{fig:sampleTraj}
\end{figure}

Using a 10-year reanalysis with this choice of $r$ and $\delta$,
we then perform twin data assimilation experiments with the ETKF,
ETKF-TC and ETKF-TV.
One realization of this experiment is shown in Figure \ref{fig:sampleTraj},
where we show $300$ analysis cycles of analyses made by the ETKF (red),
ETKF-TC (green) and ETKF-TV (blue)
for an observed variable (top panel) and
unobserved variable (middle panel),
with the black curves showing the true time series.
The lower panel shows the RMS analysis error for each method,
as averaged over all variables.
For the observed variables the ETKF-TC and ETKF-TV both outperform the standard ETKF,
which diverges after approximately 100 analysis cycles.
The ETKF-TC and ETKF-TV provide an efficient track of the truth throughout the entire duration of the experiment. 
However, while the ETKF-TV tracks the unobserved variable similarly well to the observed variable, 
the ETKF-TC diverges from the unobserved reference trajectory between analyses $175$ and $225$. 
The relative performance of the three algorithms is summarized in the time-series of the RMSE; errors are normalized using the system's natural variability $\sigma_{\rm clim}$ as above. 
The RMSE for the ETKF reaches values as large as $100\%$ of the climatological standard deviation,
meaning the assimilation process is unable to carry the information coming from the observations into the state estimate. 
However, the ETKF-TV is very successful at this task, with the RMSE converging to $25\%$ of the climatological standard deviation $\sigma_{\rm clim}$. 
The performance of the ETKF-TC is between those of the ETKF and ETKF-TV, with the RMSE converging to about $60\%$ of $\sigma_{\rm clim}$. 
Note that the RMSE here is averaged over all variables; 
in the observed subdomain the RMSE of both ETKF-TV and ETKF-TC goes well below the observational accuracy level, a sign that both filters are functioning well. 
The ETKF RMSE in the observed subdomain does not attain a similarly low level.
Such a marked difference in the skill between observed and unobserved areas for the ETKF-TC is possibly due to the use of a constant $\overline{\P}_m$. 
In the unobserved areas the analysis correction projects mostly on the span of the forecast error covariance ${\bf P}^f$,
making its correct specification crucial in those regions. 
 Note from Figure \ref{fig:pdf_dtf} that the estimated model error bias $\overline{\mathbf{b}}_m$ is approximately equal to zero;
  this means that skill improvements made by the ETKF-TC and ETKF-TV come largely from second-moment corrections in the analysis update.

\begin{figure}[h]
\centering
\includegraphics[width=\columnwidth]{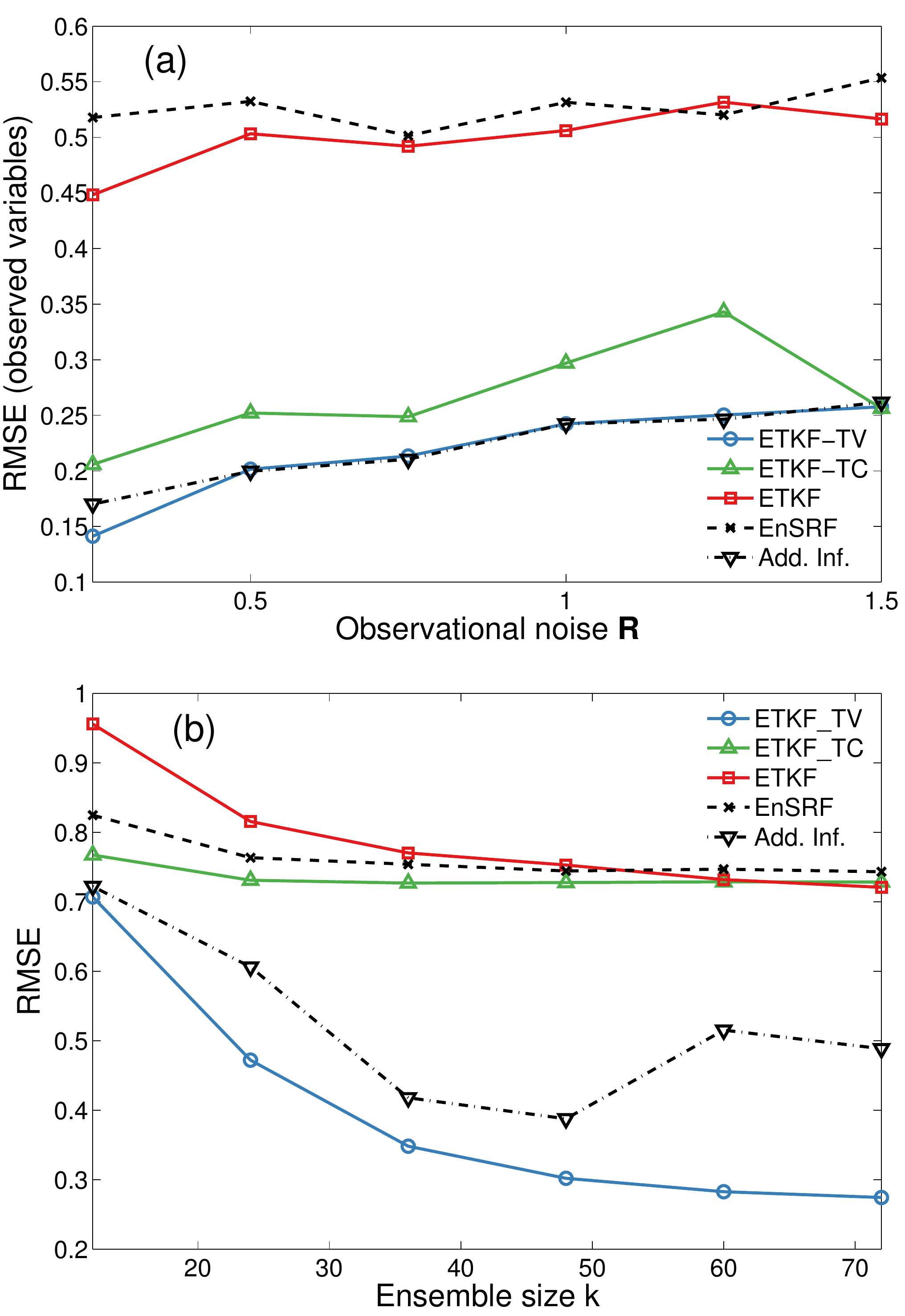}
\caption{Dependence of (a) RMSE averaged over the observed variables upon the magnitude of the observational noise $\bf R$ and (b) RMSE averaged over all variables upon the number of ensemble members for different methods.
Red: ETKF with inflation and localization;
Green: ETKF-TC;
Blue: and ETKF-TV;
Black dashed: serial EnSRF;
Black dash-dotted: ETKF with additive inflation.
Results are averaged over 200 realizations, and all errors are normalized by the climatological standard deviation $\sigma_{\rm clim}$.
}
\label{fig:errR}
\end{figure}

In Figure \ref{fig:errR} we show how the methods perform as a function of the magnitude of the observational error $\bf R$ (panel (a)) and the number of ensemble members (panel (b)).
We again set $\delta = 0.9$, $r = 3$ in all experiments and average our results over 200 realizations,
each of one month in length.
The RMS errors as a function of ensemble size are shown as averaged over all variables (panel (b)),
while for the magnitude of the observational error we focus on the RMSE averaged over the observed variables only.
The general trends observed in Figure \ref{fig:sampleTraj} are preserved in both cases,
with the ETKF-TV (blue) showing a large improvement in analysis quality over the ETKF (red),
with the ETKF-TC (green) showing a slightly more modest improvement.
The skill of each method decreases roughly monotonically as the observational error is increased,
and as the ensemble size is decreased.
Interestingly,
small ensemble sizes impact the ETKF-TV more adversely than they do the ETKF-TC,
with the latter method almost becoming preferable for $k = 12$.
The semi-independence of the ETKF-TC upon $k$ is due to the fact that the model error treatment is independent of $k$,
and the range of the matrix $\overline{\P}_m$ provides an additional subspace for error correction that is not subject to ensemble collapse.
This is an attractive property in the common situation where the ensemble size cannot be increased due to computational limitations for both the model and the ensemble-based scheme.
However,
the overall error level for the ETKF-TC remains high.
This is an indication that the model error covariance matrix is dominating the ensemble-based forecast error covariance,
and may require further tuning through the model error inflation parameter $\alpha$ (see Figure \ref{fig:errAlpha} below).


As a further comparison of the proposed model error treatments to standard techniques,
we also include in Figure \ref{fig:errR} curves corresponding to the serial ensemble square-root filter (EnSRF) (black, dashed) of \cite{Whitaker2002},
as well as an ETKF with additive inflation (black, dash-dotted) to account for model error as described by \cite{Hamill2005}.
We took the EnSRF as an example of an ensemble Kalman filter which avoids the inconsistency between analysis mean and perturbations described in Section \ref{sec:inconsistency},
and additive inflation as an example of a standard model error treatment with some similarity to the ETKF-TV.
While we refer the reader to the above articles for details of those methods,
we remark that we set $\delta = 0.9$, $r=3$ for the EnSRF as we had for the ETKF
and used our reanalysis as a proxy for a high-resolution model run in the additive inflation experiment.
The EnSRF performs comparably to the ETKF in both experiments,
while additive inflation performs as well as the ETKF-TV for the observed variables 
but less well for the unobserved variables,
leading to a higher overall RMS error in Figure \ref{fig:errR}(b).
It is not surprising that additive inflation as implemented here should perform comparably to the ETKF-TV,
as the only difference between the two methods is in the sampling of the dataset of reanalysis increments to produce $\et_{i,j}$ in (\ref{TV}).
Following \cite{Hamill2005},
in the additive inflation experiment the sampling is done directly from the record of reanalysis increments,
while in the ETKF-TV this is done from the multivariate covariance matrix $\P_m$.
While this is expected to lead to significant differences when the model error distribution is far from Gaussian,
from Figure \ref{fig:cov_dtf} we know that this is not the case here and the skill of additive inflation and ETKF-TV are comparable.

\begin{figure}[htbp]
\centering
\includegraphics[width=\columnwidth]{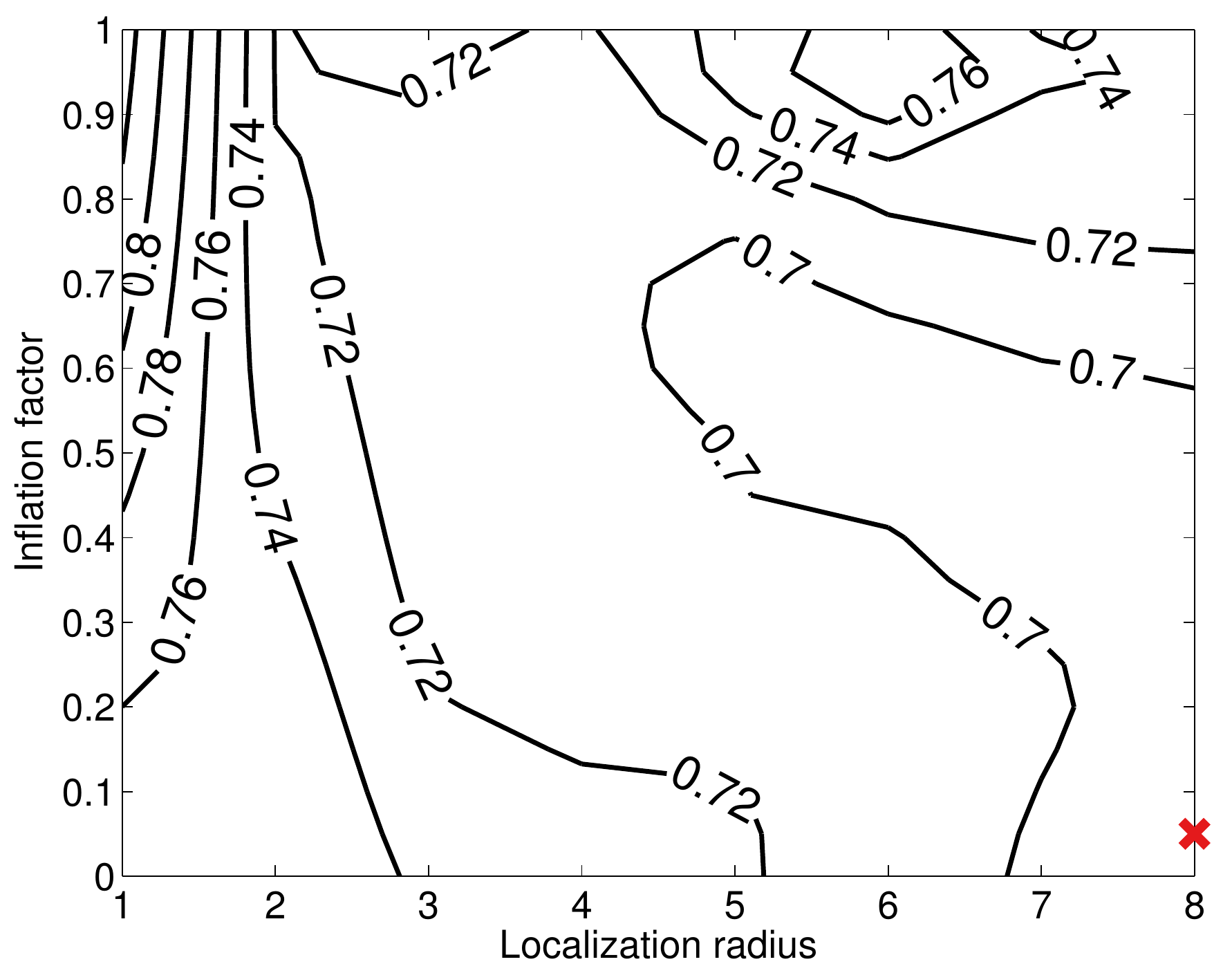}\\
\includegraphics[width=\columnwidth]{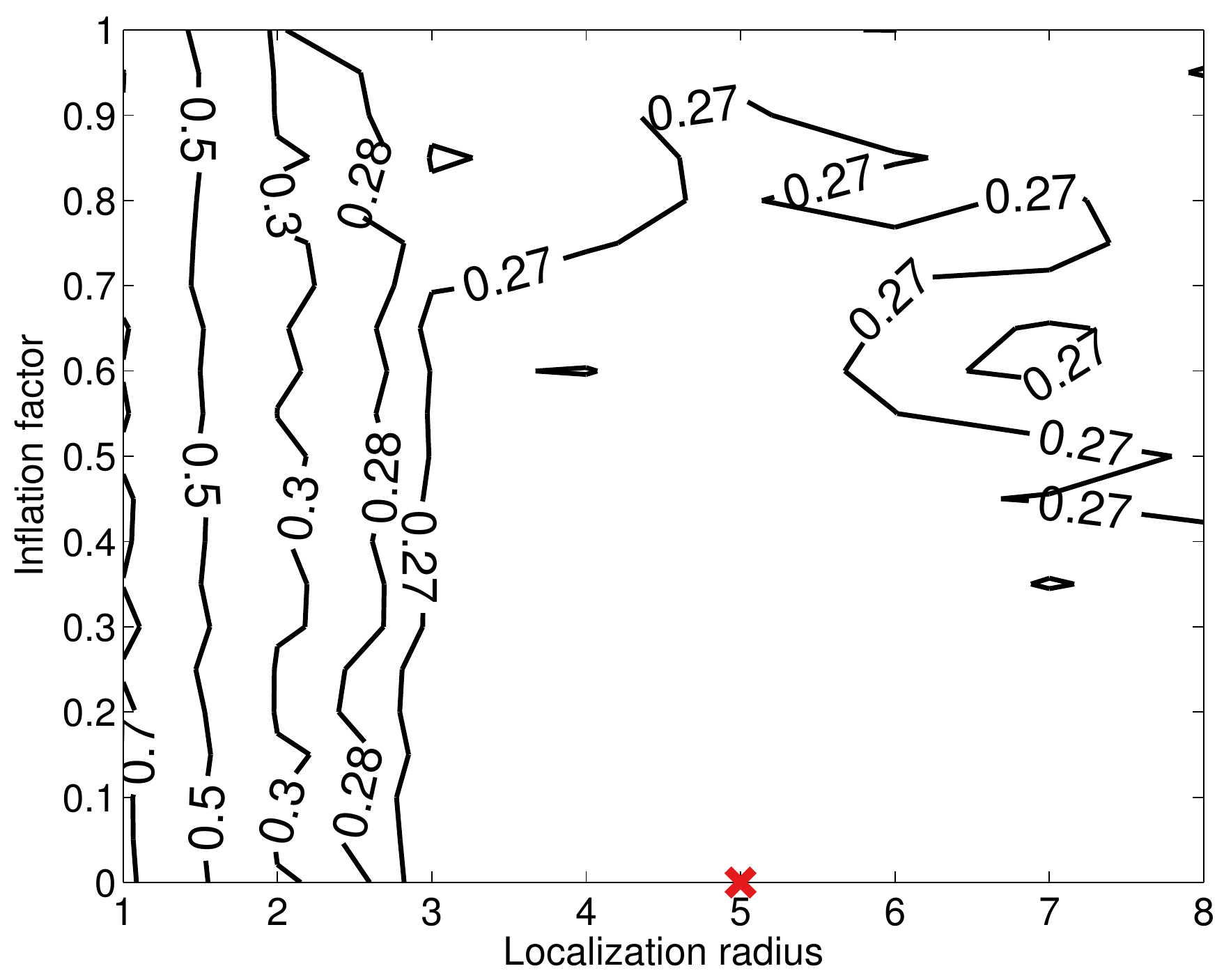}
\caption{Contour plot of RMS errors for the ETKF-TC (top) and ETKF-TV (bottom) as a function of inflation factor $\delta$ and localization radius $r$. The red crosses show the location of the minima. RMS errors are normalized by the climatology standard deviation $\sigma_{\rm clim}$, and are averaged over 200 realizations of one month in duration.}\label{fig:inflationLocalisationCovStoch}
\end{figure}

We now investigate whether our model error treatments can make further improvements over the standard ETKF by re-tuning the inflation and localization parameters $\delta$ and $r$ in the experiment stage.
To do this we use the same reanalysis data set with $\delta = 0.9$ and $r=3$ to build the model error biases (\ref{bias_TC-2}) and covariance matrices (\ref{cov_aprx_3}),
however vary the $\delta$ and $r$ used in the ETKF-TC and ETKF-TV experiments.
The contour plots in Figure \ref{fig:inflationLocalisationCovStoch} show the results of these experiments for the ETKF-TC (top panel) and ETKF-TV (bottom panel),
displaying the RMS errors over all variables and as averaged over 20 realizations each.
Firstly,
the ETKF-TV produces the best analyses for all values of $\delta$ and $r$,
showing RMS errors approximately half those of the ETKF-TC for $r > 1$.
Secondly,
the two methods are optimized by slightly different values of $\delta$ and $r$.
The ETKF-TC obtains its best performance when $\delta =0.05$ and $r = 8$,
while the ETKF-TV is optimized when $\delta = 0$ and $r=5$ (approximately).
This implies that the ETKF-TC only underestimates the actual forecast error variance by approximately $5\%$,
and it is essentially correctly estimated by the ETKF-TV 
(compared with approximately $90\%$ underestimation for the ETKF).
Moreover, the large values found for the optimal localization radius $r$ indicate that both model error treatments have essentially mitigated the misrepresentation of the long-range spatial correlations by spurious ensemble effects, 
compared to the standard ETKF.  
Both methods outperform the optimized standard ETKF, 
with RMSE equal to 0.256 for the ETKF-TV with optimal tuning and 0.690 for the ETKF-TC with optimal tuning compared with
0.709 for the standard ETKF with optimal tuning.
Interestingly, the skill dependence of ETKF-TC and ETKF-TV on $\delta$ and $r$ is very similar and clearly different from that of the standard ETKF. 
From Figure \ref{fig:inflationLocalisationCovStoch} we observe that for $r\le3$ the performance of the two methods depends only on the localization factor. 
For larger $r$ the skill of either method almost converges,
with only minor dependence on $\delta$ and $r$. 
Both the ETKF-TC and ETKF-TV therefore appear very stable with respect to the tunable parameters $\delta$ and $r$, 
an attractive property when considering their implementation with large numerical systems.

Following \cite{Carrassi2011}, we investigate the effect of varying the model error inflation factor $\alpha$ in (\ref{ETKFTC_b})-(\ref{TV}).
As was discovered in \cite{Carrassi2011},
we again find here that some deflation of the model error covariance matrix $\overline{\P}_m$ used in both the ETKF-TC and ETKF-TV can actually be beneficial to those methods.
In Figure \ref{fig:errAlpha} we show how the RMS analysis errors in ETKF-TC depend upon the model error inflation/deflation factor $\alpha$.
Each point is averaged over 200 realizations,
and shows that a deflation of the model error term through setting $\alpha = 0.2$ improves the performance of the ETKF-TC by approximately $40\%$,
while setting $\alpha = 0.5$ improves the ETKF-TV very slightly (however this improvement is not statistically significant).
That the best results are obtained for $\alpha<1$ and that the RMS errors worsen for $\alpha > 1$ implies that the approximations (\ref{bias_TC-2}) and (\ref{cov_aprx_3}) are actually overestimations of the actual bias and covariance.
This is because (\ref{bias_TC-2}) and (\ref{cov_aprx_3}) incorporate some initial condition error into the estimation of $\overline{\P}_m$,
so optimal results are obtained when $\alpha$ is tuned accordingly.
The response to $\alpha$ for the two algorithms is however very different, 
with the ETKF-TC showing the largest dependence. 
This is not surprising given the time-constant model error treatment of the ETKF-TC,
and is also consistent with the discussion in relation with Figure 6(b).
The ETKF-TV depends less on $\alpha$ because different optimal model error sizes may be required at each analysis step.

It is worth studying the relative performance of the algorithms for different levels of model error. 
This is controlled in the Lorenz-96 model by modifying the parameter $h$. 
However this factor has also a strong interplay with other dynamical aspects of the model, 
such the spectrum of the Lyapunov exponents and the amplitude of the dominant Lyapunov exponent $\gamma_{\rm max}$. 
As such, it is difficult to draw a clear relationship between $h$ and the performance of our two model error treatments and so we cannot expect the relative performance of ETKF-TC/TV over the ETKF to improve monotonically with $h$.
In Figure \ref{fig:errh} we show the ratio of the ETKF error to that of the ETKF-TC and ETKF-TV,
so that values greater than 1 represent an improvement in analysis skill.
Figure \ref{fig:errh} shows that a monotonic trend is almost found for $h\leq1$, as the model error decreases. 
For larger $h$ the curves resemble the dependence of $\gamma_{\rm max}$ over $h$ (see the inset of Figure \ref{fig:LE_h}),
suggesting that in a stabler model configuration the gain in skill of the ETKF-TC/TV is smaller than in more unstable cases.

\begin{figure}[htbp]
\centering
\includegraphics[width=\columnwidth]{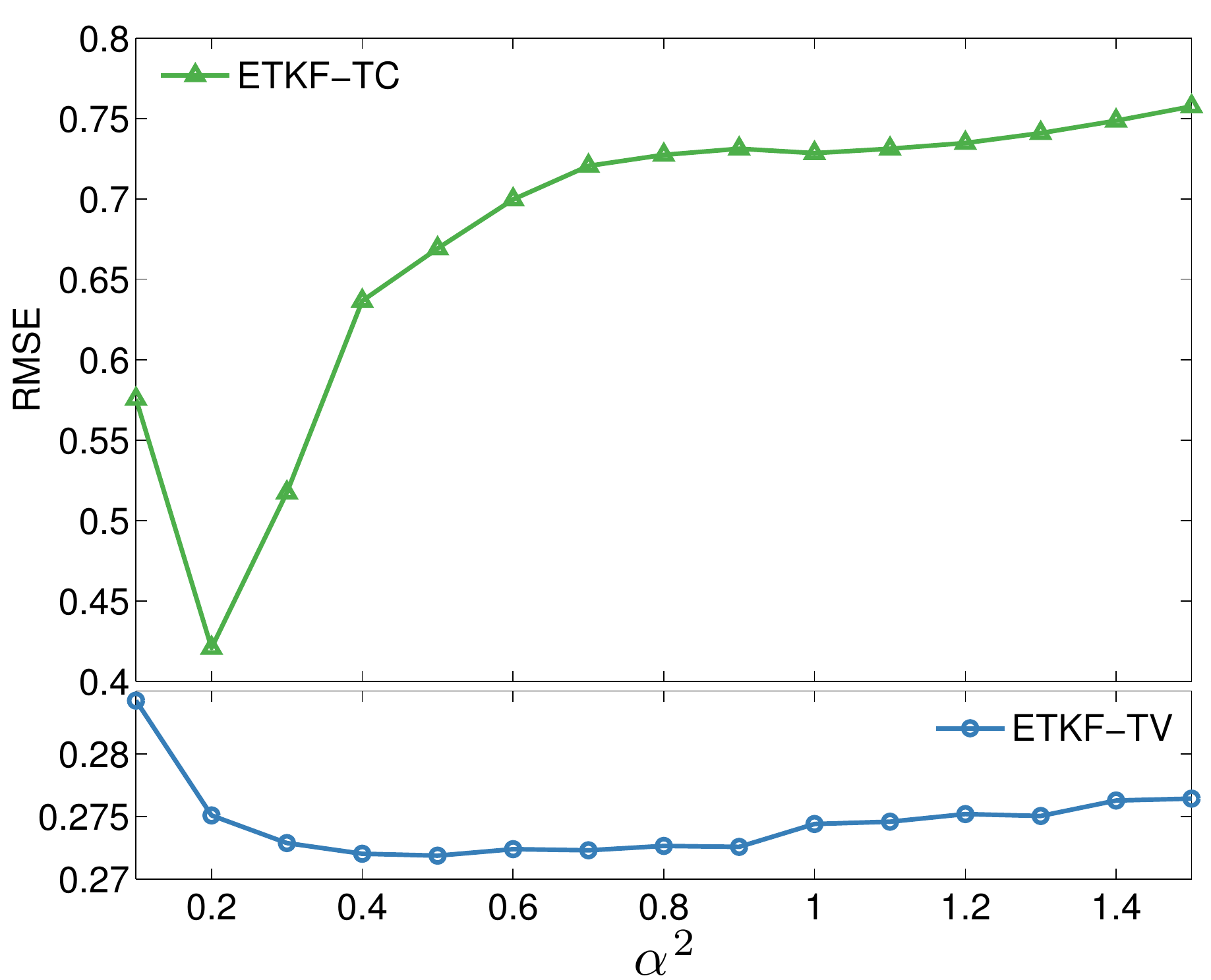}
\caption{Dependence of RMS errors for the ETKF-TC upon the magnitude of the model error inflation factor $\alpha$.
}
\label{fig:errAlpha}
\end{figure}

\begin{figure}[htbp]
\centering
\includegraphics[width=\columnwidth]{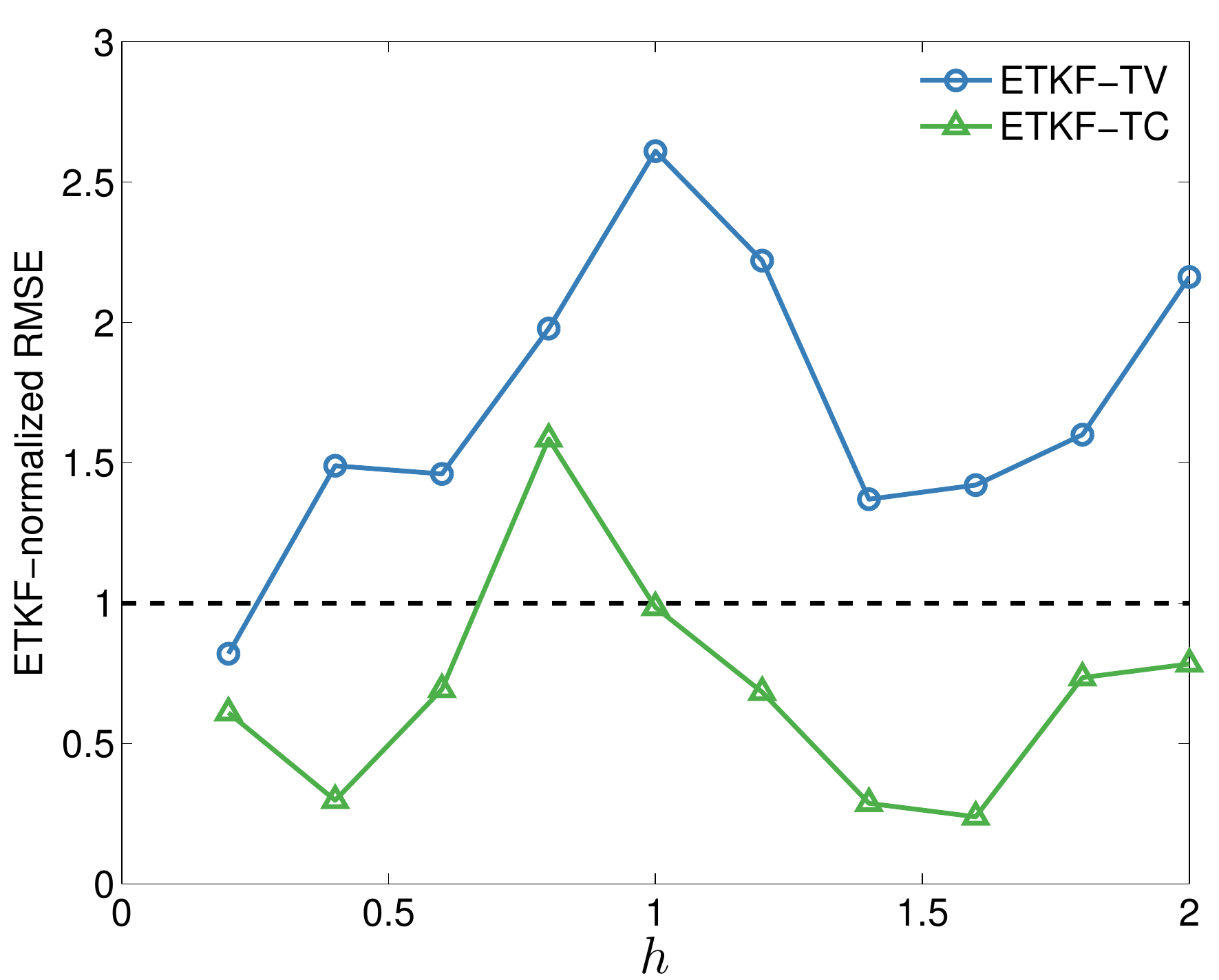}
\caption{Dependence of the skill of ETKF-TC and ETKF-TV upon the magnitude of the size of the model error $h$.
We show the ratio of the ETKF errors divided by each method, so that values greater than 1 represent a skill improvement.
All values are averaged over 200 realizations.
}
\label{fig:errh}
\end{figure}

\section{Discussion\label{sec:discussion}}

Following the formulation of \cite{Nicolis2004}
we have derived a method for the treatment of time-correlated model error in the short-time limit for an ensemble Kalman filter.
In the absence of complete knowledge of the true atmospheric state,
we used a database of historical reanalysis increments to approximate the model error biases and covariance matrices required in the least-square based analysis update.
One clear drawback of this choice is that its accuracy depends on the properties of the observational network, so that in unobserved or poorly-observed areas the analysis increments will be almost zero even in the presence of model error. 
This dependency is however less dramatic than in the closely related approach of using the innovations 
(differences between observation and forecast)
to estimate or tune background or observational error covariance matrix (see {\it e.g.} \cite{Desroziers2005}).
In this latter case in fact only the projection of the error covariance matrix on the observational space can be estimated.
The statistics of the analysis increments have been also used by \cite{Cullen2010} to estimate the so-called ``regularization matrix'' (see also \cite{Cullen2010a}),
a newly introduced term used to improve the fit to the observations.
In this view, the regularization term is intended to account for all the system error 
(also connected to the assimilation procedure itself) 
and not just the error originating from the underlying model dynamics. 

We studied two potential implementations of this method for an ETKF,
namely the ETKF-TC which assumed the model error correction term to be constant in time
and the ETKF-TV where the model error correction was allowed to vary randomly after each forecast.
In the latter case, 
the noise is sampled from the assumed model error statistics and its contribution to the forecast error is propagated forward using a short-time approximation. 
These methods were compared with the standard approach for dealing with model error in the ETKF of tuning the inflation factor and localization radius empirically to minimize RMS errors
through numerical experiments on the Lorenz-96 model.

Both the ETKF-TC and ETKF-TV showed improvements upon the ETKF with optimized inflation and localization parameters,
with the ETKF-TV showing the lowest RMS errors of all three methods.
The two methods with deterministic model error treatment showed further improvements once inflation and localization were retuned,
but importantly were found to be less sensitive to the precise tuning of these parameters than the ETKF was.
This is a desirable property when considering the potential implementation of ETKF-TC or ETKF-TV in an operational setting.
Room for additional improvements in the ETKF-TC and -TV are based on the recognition that the adopted model error treatment is prone to suffer from an overestimation of the actual model error,
due to the presence of initial condition error. 
Results confirmed this conjecture and showed that the skill of both ETKF-TC and -TV can be further improved by deflating the associated model error contribution to the state-estimation error. 
This is particularly relevant for the ETKF-TC in view of its description of the model error being constant in time,
which has both advantages and disadvantages.
A stationary model error treatment suggests that the ETKF-TC should only be expected to perform well on average,
and indeed we did find that the ETKF-TV was better able to capture intermittent events such as are expected in forecasting on a meteorological time scale.
However, the independence over the ensemble size of the model error contribution in the ETKF-TC
 makes this algorithm less sensitive to errors arising from small ensemble size effects than the time-varying ETKF-TV was.
The overall performance of the ETKF-TV is markedly better than the ETKF-TC, however the aforementioned features make the ETKF-TC attractive in situations where a pre-existing ensemble scheme is already implemented and the computational recourses are already fully exploited.

This study has been carried out using a simplified chaotic dynamical system that has allowed for an extensive and robust exploration of the methods under consideration.
Furthermore, the use of this model has enabled us to use the ETKF in all numerical experiments,
which is known to not be practical in operational contexts due to the inability to store the forecast covariance matrix $\P^f$.
An important step forward will be through applications of the same approach to model and observational setups of increasing complexity.
In parallel to this application-oriented research direction, 
the authors are studying alternative formulations of the square-root filter in which the model error treatment does not give rise to the inconsistency described in Section 4.3.
It may also be possible to use methods such as the LETKF
or EnSRF which allow for localization to be implemented to the forecast perturbations.
How such filters perform with the proposed model error treatment is a direction for future research,
and how such methods compare with standard procedures is of great relevance in the ongoing debate over model error and ensemble filtering.

A logical further extension of this work would be to explicitly incorporate the temporal evolution of model error.
Model error temporal correlations are allowed in a weak-constraint variational framework \citep{Tremolet2006},
and the use of a deterministic method for their estimation has been introduced in \cite{Carrassi2010}. 
An extension of the present study along this same line implies the use of ensemble-based smoother, a research direction that we leave for future work.  

\ack
The authors thank Patrick Raanes for helpful discussions,
as well as the three reviewers for helping us to improve this manuscript.
LM was supported by NSF grant DMS-0940271, and is grateful for the computational resources provided by the Vermont Advanced Computing Core which is supported by NASA (NNX 08A096G), and the Vermont Complex Systems Center. 
AC was financed through the IEF Marie Curie Project INCLIDA of the FP7,
as well as EU-FP7 project SANGOMA under grant agreement no. 283580.

\bibliographystyle{wileyqj}
\bibliography{library}

\clearpage

\end{document}